\documentclass[letterpaper]{article}

\usepackage[T1]{fontenc}

\usepackage{geometry}
\usepackage{setspace}

\usepackage[style = chem-acs]{biblatex}
\usepackage{graphicx}
\usepackage{caption}
\usepackage{float}
\newfloat{scheme}{htbp}{los}
\floatname{scheme}{Scheme}
\floatname{chart}{Chart}
\newfloat{graph}{htbp}{loh}

\usepackage{chemformula} 
\usepackage[version = 4]{mhchem} 

\usepackage{authblk}
\author[1,2]{Bo Liang$^\dagger$}
\author[1,3]{Xue Li$^\dagger$}
\author[4]{Congcong Le$^\dagger$*}
\author[1,2]{Zirui Wu$^\dagger$}
\author[1,2]{Wenpei Zhu}
\author[1,2]{Neng Cai}
\author[1,2]{Yong-Chang Lau}
\author[5]{Xianxin Wu}
\author[6]{Jiayu Liu}
\author[7]{Zhanfeng Liu}
\author[7]{Hongen Zhu}
\author[7]{Tongrui Li}
\author[7]{Zhicheng Jiang}
\author[6]{Yu Huang}
\author[6]{Wenchuan Jing}
\author[6]{Xun Ma}
\author[8]{Qi Jiang}
\author[9]{Hang Li}
\author[1,2]{Zhihao Cai}
\author[10]{Xuezhi Chen}
\author[1,2]{Gexing Qu}
\author[10]{Yiwei Cheng}
\author[1,2]{Bing-Jie Chen}
\author[8]{Zhengtai Liu}
\author[7]{Dawei Shen}
\author[8]{Mao Ye}
\author[7]{Shengtao Cui}
\author[7]{Zhe Sun}
\author[11]{Koji Miyamoto}
\author[11]{Taichi Okuda}
\author[11]{Kenya Shimada}
\author[8]{Yaobo Huang}
\author[8]{Zhenhua Chen}
\author[1,2,12]{Lin Zhao}
\author[1,2]{Baojie Feng}
\author[1,2]{Xinguo Ren*}
\author[9]{Wenhong Wang*}
\author[1,2,12]{Xingjiang Zhou*}
\author[1,2,12]{Guodong Liu*}
\affil[1]{Beijing National Laboratory for Condensed Matter Physics, Institute of Physics, Chinese Academy of Sciences, Beijing 100190, China}
\affil[2]{School of Physical Sciences, University of Chinese Academy of Sciences, Beijing 100049, China}
\affil[3]{High-tech Institute, FanGong-ting South Street on the 12$^{\mathrm{th}}$, Qingzhou, Shandong, China}
\affil[4]{Hefei National Laboratory, Hefei 230088, China}
\affil[5]{CAS Key Laboratory of Theoretical Physics, Institute of Theoretical Physics, Chinese Academy of Sciences, Beijing, China}
\affil[6]{Shanghai Institute of Microsystem and Information Technology, Chinese Academy of Sciences, Shanghai 200050, China}
\affil[7]{National Synchrotron Radiation Laboratory and School of Nuclear Science and Technology, University of Science and Technology of China, Hefei 230026, China}
\affil[8]{Shanghai Synchrotron Radiation Facility, Shanghai Advanced Research Institute, Chinese Academy of Sciences, Shanghai 201210, China}
\affil[9]{Tiangong University, Tianjin 300387, China}
\affil[10]{Shanghai Institute of Applied Physics, Chinese Academy of Sciences, Shanghai 201800, China}
\affil[11]{Research Institute for Synchrotron Radiation Science,
Hiroshima University, 739-0046 Higashi-Hiroshima, Japan}
\affil[12]{Songshan Lake Materials Laboratory, Dongguan, Guangdong 523808, China}

\title{Termination-Dependent Surface States and Magnetic Fingerprints of Chiral Helimagnet Cr$_{1/3}$TaS$_2$}
\date{*Email: lecongcong@hfnl.cn, renxg@iphy.ac.cn, wenhongwang@tiangong.edu.cn, xjzhou@iphy.ac.cn, gdliu\_arpes@iphy.ac.cn}

\begin{document}

\maketitle

\begin{abstract}
  Chiral helimagnets based on intercalated transition-metal dichalcogenides, characterized by nano-scale spin ordering, provide a powerful route to engineer chiral spin textures (\textit{e.g.}, the topologically protected magnetic solitons) and emergent electronic functionality at reduced dimensions, where surface and interface states often dominate device operation. However, despite growing interest, direct experimental studies of termination-dependent surface electronic structures and their temperature-driven magnetic evolution remain largely unexplored, hindering a microscopic understanding of the electronic states that is crucial for the development of low-dimensional spintronic devices. Here, for the first time, taking Cr$_{1/3}$TaS$_2$ as a representative example, we systematically investigate the termination-dependent surface electronic states of the chiral helimagnets and uncover their distinct temperature evolution across the magnetic transition ($T_C$$\sim$142\,K) by combining high-resolution angle-resolved photoemission spectroscopy with a micro-focused beam and surface-state-resolved first-principles calculations. The TaS$_2$-terminated surface hosts folded monolayer-like TaS$_2$ bands under the $\sqrt{3}\,\times\sqrt{3}$ superlattice potential and a shallow triangular electron pocket at the superlattice $\bar{K}$ point arising from Cr--Ta orbital hybridization. In contrast, the Cr-terminated surface exhibits reconstructed hole pockets with pronounced magnetic band splitting. This splitting disappears above $T_C$ and closely follows the chiral helimagnetic order parameter, providing a direct spectroscopic fingerprint of chiral helimagnetic order. In addition, multiple ultranarrow Cr-$d$-derived surface flat bands are resolved. These findings establish Cr$_{1/3}$TaS$_2$ as a model system in which surface electronic states are strongly coupled to chiral magnetism, opening new opportunities for chiral spintronic and valleytronic micro/nanodevices.
\end{abstract}

\section*{Keywords}

angle-resolved photoemission spectroscopy, chiral helimagnetism, magnetic solitons, surface electronic states, magnetic band splitting, intercalated transition-metal dichalcogenides

\section{Introduction}

Layered transition-metal dichalcogenides (TMDs) have emerged as a central platform for both fundamental research and device applications in two-dimensional materials. Magnetic intercalation in TMDs provides an effective route to engineer unconventional magnetic orders and emergent electronic phenomena in low-dimensional materials. By inserting magnetic transition-metal ions (V, Cr, Mn, Fe, Co, Ni) into the van der Waals gaps, the crystal symmetry, exchange interactions, and spin--orbit coupling (SOC) can be simultaneously tuned, leading to a variety of exotic magnetic phases, including altermagnetism\autocite{CoNb4Se82025} and non-coplanar antiferromagnetic order.\autocite{takagi2023spontaneous,park2023tetrahedral} In particular, chromium (Cr)-intercalated TMDs such as Cr$_{1/3}$NbS$_2$ and Cr$_{1/3}$TaS$_2$ have attracted considerable attention, as the ordered Cr sublattice breaks inversion symmetry, endowing the system with non-centrosymmetric crystal symmetry and structural chirality, and stabilizes a nano-scale chiral helimagnetic order (CHM) as the magnetic ground state.\autocite{MORIYA1982209,TogawaPRL2012,CHMCTS} This long-period helical spin order originates from the competition between ferromagnetic (FM) exchange interactions and the antisymmetric Dzyaloshinskii--Moriya (DM) interaction\autocite{DZYALOSHINSKY1958241,Moriya} enabled by SOC in a non-centrosymmetric lattice.

Upon applying an in-plane magnetic field, the CHM continuously evolves into a chiral soliton lattice (CSL), consisting of a one-dimensional (1D) periodic array of topologically protected nano-scale spin solitons.\autocite{TogawaPRL2012,Togawa2017PhysRevB} These particle-like 1D spin textures can be viewed as 1D counterparts of two-dimensional (2D) and three-dimensional (3D) skyrmions, inheriting non-trivial real-space spin topology.\autocite{Braun01022012} As a consequence, chiral magnetic solitons exhibit remarkable stability and give rise to a range of unconventional transport phenomena, including negative magnetoresistance,\autocite{NegativeMRPRL} quantized magnetoresistance steps,\autocite{TogawaPRB,ThicknessControllPRL} giant planar Hall effect,\autocite{PhysRevResearch.4.013134} and topological Hall effect.\autocite{THE} These observations highlight the strong coupling between chiral magnetism and electronic properties in intercalated TMDs. The ability to host such topologically driven transport anomalies, combined with the fact that in ultrathin exfoliated crystals the number of magnetic solitons can be tailored by selecting an appropriate crystal thickness,\autocite{ThicknessControllPRL} and that individual solitons can be created and annihilated in a controlled manner by external magnetic fields,\autocite{TogawaPRB} makes Cr-intercalated chiral helimagnets an exciting platform for both fundamental studies and spintronic applications.

Among this family, Cr$_{1/3}$TaS$_2$ stands out as a particularly promising chiral helimagnet owing to the incorporation of the heavy 5d element Ta, which significantly enhances the SOC compared to its Nb-based counterparts. The strengthened SOC amplifies the DM interaction, leading to modified magnetic energy scales and a reduced helical pitch. Consistent with this, Cr$_{1/3}$TaS$_2$ exhibits a higher Curie temperature ($T_C$) and a larger critical magnetic field ($H_C$) than Cr$_{1/3}$NbS$_2$,\autocite{XieLL2022} indicative of a more robust chiral magnetic ground state. Correspondingly, the helical period of the CHM ground state in Cr$_{1/3}$TaS$_2$ is nearly halved compared to that in Cr$_{1/3}$NbS$_2$,\autocite{CTSperiod,TogawaPRL2012} placing it at a substantially shorter length scale that enables a higher density of chiral magnetic solitons, thereby facilitating device miniaturization. From an application-oriented perspective, the combination of a relatively high transition temperature and a high density of topologically protected solitons establishes Cr$_{1/3}$TaS$_2$ as an ideal candidate for low-power spintronic devices, where chiral magnetic solitons can serve as mobile information carriers driven by ultralow current densities.\autocite{kishine2010sliding,tokushuku2017tunable,laliena2020dynamics,Masell2021,osorio2022response}

Despite extensive progress in elucidating the magnetic properties of chiral helimagnets, their genuine electronic structure of chiral helimagnets---particularly at surfaces and across magnetic phase transitions---remains largely unexplored. Given that the macroscopic physical properties of quantum materials are fundamentally governed by their low-energy electronic structure near the Fermi level ($E_F$), and that angle-resolved photoemission spectroscopy (ARPES) provides the only direct momentum-resolved experimental probe of electronic band dispersions, a comprehensive ARPES investigation is therefore indispensable. However, most previous studies have primarily interpreted experimental observations from the perspective of bulk electronic structure, without explicitly considering the role of surface states in spectroscopic measurements.\autocite{CNS2016PhysRevB,cNS2020CP,CYLCNSPRB} As a result, it remains unclear how surface termination and near-surface magnetic correlations reshape the low-energy electronic states that are directly relevant to device interfaces.

Although a recent study has noted indications of distinct surface terminations in Cr-intercalated chiral helimagnets,\autocite{XieLLARPES} the existing results remain controversial. In particular, the observed band features have not been unambiguously correlated with specific surface terminations, and the experimental spectra were interpreted mainly through comparisons with bulk band-structure calculations, without employing explicit surface-state electronic-structure calculations for reliable band assignment. Moreover, the absence of systematic temperature-dependent studies in different terminations has hindered a clear understanding of the interplay between surface electronic states and magnetism. This limitation is especially critical for van der Waals layered materials, as practical spintronic and electronic devices inevitably operate in ultrathin films, where surface and interface electronic states can dominate charge transport and spin-dependent functionalities.

In addition, substantial controversy persists regarding the reported magnetically split bands in Cr$_{1/3}$NbS$_2$. Sirica \textit{et al.}\autocite{cNS2020CP} reported a splitting of the outermost hole-like Fermi pocket centered at the $\bar{\Gamma}$ point and, based on resonant photoemission spectroscopy (ResPES) observations of Cr-derived spectral weight at the Fermi level, proposed the presence of a strong exchange (Hund’s) coupling between itinerant electrons and localized Cr moments in Cr$_{1/3}$NbS$_2$. By contrast, Qin \textit{et al.}\autocite{CYLCNSPRB} reported that the splitting occurs in the innermost hole-like Fermi pocket near $\bar{\Gamma}$ and persists well above the Curie temperature, pointing to short-range magnetic order beyond a simple Stoner-type itinerant picture. Notably, neither study explicitly considered the possible influence of distinct surface electronic states in chiral helimagnets, and both works interpreted the observed band splittings within a purely bulk framework, potentially overlooking surface-derived contributions that actually dominate the measured electronic structure.

In this work, for the first time, we combine high-resolution micro-ARPES and first-principles calculations to systematically investigate the termination-dependent electronic structure of Cr$_{1/3}$TaS$_2$ and its temperature evolution across the CHM transition. Our results provide the first direct insight into how the surface electronic structure reflects the underlying magnetic properties, revealing marked differences between TaS$_2$- and Cr-terminated surfaces and how these surface states evolve in response to changes in temperature. These findings not only shed light on the microscopic electronic origins of chiral magnetism in Cr$_{1/3}$TaS$_2$, but also help clarify the genuine electronic structure of chiral helimagnetic materials more broadly, while offering guidance for the rational design of high-performance low-dimensional spintronic devices based on chiral magnetic solitons.

\section{Results and discussion}

\subsection{Crystal structure, magnetic properties, and bulk electronic structure of Cr$_{1/3}$TaS$_2$}

\begin{figure}
  \includegraphics[width=\linewidth]{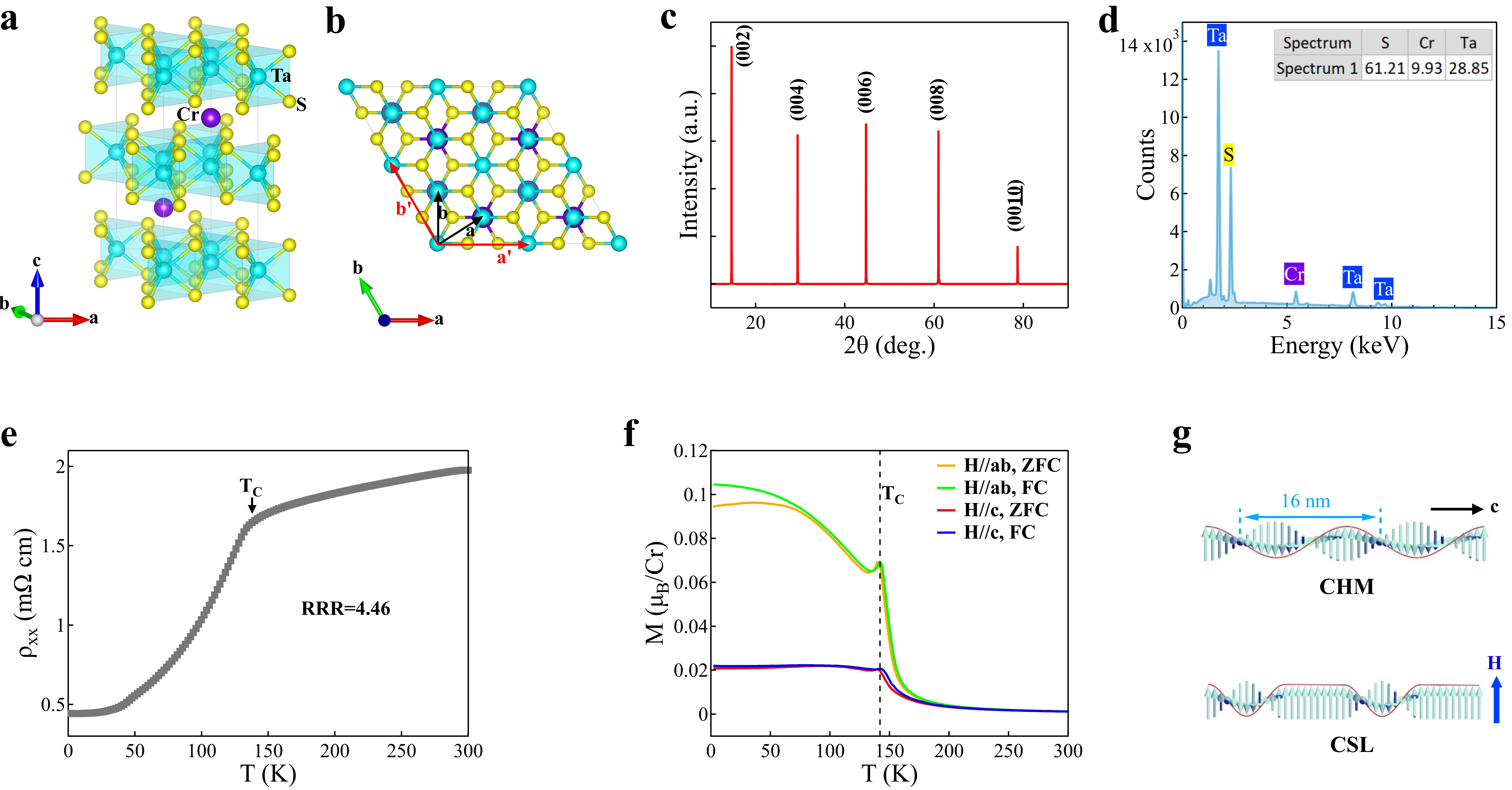}
  \caption{Crystal structure and magnetic properties of Cr$_{1/3}$TaS$_2$. (a) Crystal structure of Cr$_{1/3}$TaS$_2$, where Cr atoms are intercalated between adjacent TaS$_2$ layers. (b) Top view of the Cr$_{1/3}$TaS$_2$ lattice along the $c$-axis, illustrating the in-plane ordering of intercalated Cr atoms. (c) X-ray diffraction (XRD) pattern of a high-quality Cr$_{1/3}$TaS$_2$ sample. (d) EDX spectrum of a Cr$_{1/3}$TaS$_2$ single crystal, showing characteristic peaks from Ta, S, and Cr. (e) Temperature dependence of the longitudinal resistivity $\rho_{xx}$ measured at zero magnetic field, with a residual resistivity ratio (RRR) of 4.46. (f) Temperature-dependent magnetization measured under zero-field-cooling (ZFC) and field-cooling (FC) conditions, with magnetic fields applied within the $ab$ plane and along the $c$ axis, respectively. (g) Schematic illustration of the chiral magnetic structure below $T_C$: a chiral helical magnet (CHM) at zero magnetic field and a chiral soliton lattice (CSL) stabilized under a magnetic field applied perpendicular to the $c$ axis. }
  \label{Lattice}
\end{figure}

Cr$_{1/3}$TaS$_2$ crystallizes in a well-ordered structure with Cr intercalated at a one-third concentration, in which Cr atoms occupy the octahedral sites between adjacent TaS$_2$ layers, forming a long-range $\sqrt{3}\,\times\sqrt{3}$ superlattice (Figure \ref{Lattice}a,b). X-ray diffraction (XRD) measurements confirm the high crystalline quality and the phase purity of the synthesized single crystals (Figure \ref{Lattice}c), while energy-dispersive X-ray spectroscopy (EDX) further verifies the elemental ratio consistent with the nominal composition (Figure \ref{Lattice}d). The in-plane electrical resistivity exhibits metallic behavior over the entire measured temperature range from 2\,K to 300\,K, with a pronounced anomaly marking the magnetic transition (Figure \ref{Lattice}e). Magnetization measurements ($M-T$) reveal strong anisotropy between in-plane ($H \parallel ab$) and out-of-plane ($H \parallel c$) field orientations (Figure \ref{Lattice}f), indicative of a highly anisotropic magnetic ground state. Below the Curie temperature ($T_C$$\sim$142\,K), Cr$_{1/3}$TaS$_2$ enters a chiral magnetic regime, in which a CHM is stabilized at zero field and can be continuously driven toward a CSL under an in-plane magnetic field ($H \parallel ab$), as schematically illustrated in Figure \ref{Lattice}g.

\begin{figure}
  \includegraphics[width=\linewidth]{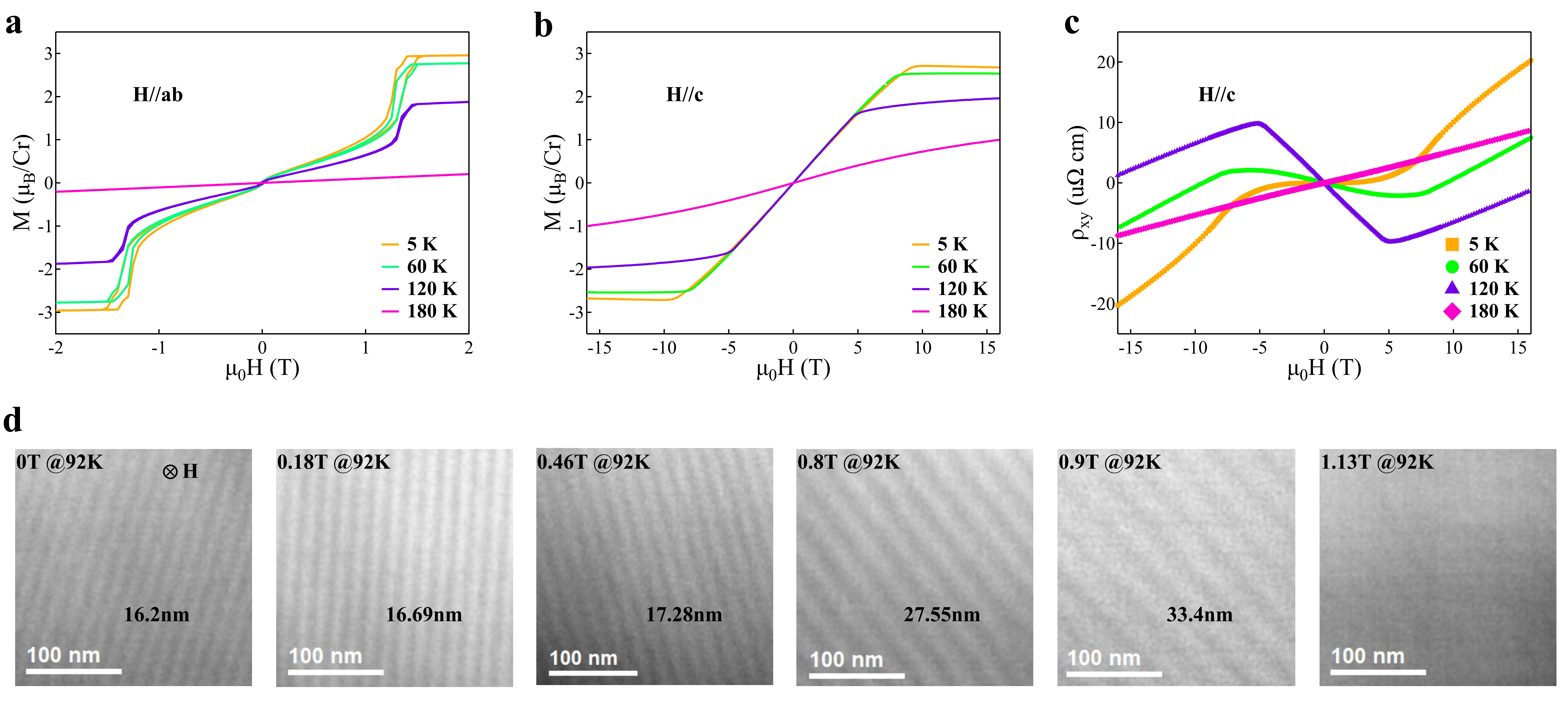}
  \caption{Field-induced evolution of magnetic properties in Cr$_{1/3}$TaS$_2$. (a) Magnetic field dependence of the magnetization ($M$-$H$) measured with the magnetic field applied within the $ab$ plane ($H\,\,||\,\,ab$) at 5, 60, 120, and 180\,K. (b) $M$-$H$ curves measured with the magnetic field applied along the crystallographic $c$ axis ($H\,\,||\,\,c$) at the same temperatures in (a). (c) Temperature dependence of the Hall resistivity $\rho_{xy}$ as a function of magnetic field ($H\,\,||\,\,c$) from 5--180\,K for Cr$_{1/3}$TaS$_2$ single crystal. (d) Lorentz-TEM images acquired at 92\,K, showing the continuous evolution of chiral helical magnetic structures under in-plane magnetic field ranging from 0\,T to 1.13\,T.}
  \label{TEM}
\end{figure}

To further elucidate the magnetic anisotropy and field response, detailed field-dependent magnetization ($M-H$) measurements were performed with magnetic fields applied along the in-plane ($H \parallel ab$) and out-of-plane ($H \parallel c$) directions (Figure \ref{TEM}a,b). The markedly different saturation fields unambiguously identify the in-plane direction as the easy magnetization plane. At 5\,K, the saturation field for $H \parallel ab$ is approximately 1.52\,T, which is about 34 times larger than that of Cr$_{1/3}$NbS$_2$ single crystals,\autocite{li2022angular} while the saturation field along the $c$ axis reaches $\sim$9.1 T, exceeding that of Cr$_{1/3}$NbS$_2$ by a factor of $\sim$5.7. Notably, as shown in Figure \ref{TEM}a, the pronounced magnetic hysteresis emerges at low temperatures when an in-plane magnetic field is applied, and the corresponding coercive field increases systematically upon cooling---from $\sim$0.05\,T at 60\,K to $\sim$0.1\,T at 5\,K. Such a cooling-enhanced coercive field has been previously attributed to the nucleation and annihilation processes of the CSL, which require overcoming distinct energy barriers associated with the topologically nontrivial spin textures.\autocite{CHMCTS,clements2017critical} Correspondingly, Hall resistivity measurements reveal a sizable anomalous Hall contribution at low temperatures (Figure \ref{TEM}c), consistent with a strong SOC-induced Berry curvature associated with the magnetic order. Direct real-space evidence (Figure \ref{TEM}d) for the CHM ground state below $T_C$ and for its continuous field-driven evolution into a CSL is provided by Lorentz transmission electron microscopy (Lorentz-TEM), which visualizes the field-driven crossover from a low-field CHM with a characteristic period of $\sim$16\,nm to a CSL at higher in-plane magnetic fields ($H \parallel ab$).

\begin{figure}
  \includegraphics[width=\linewidth]{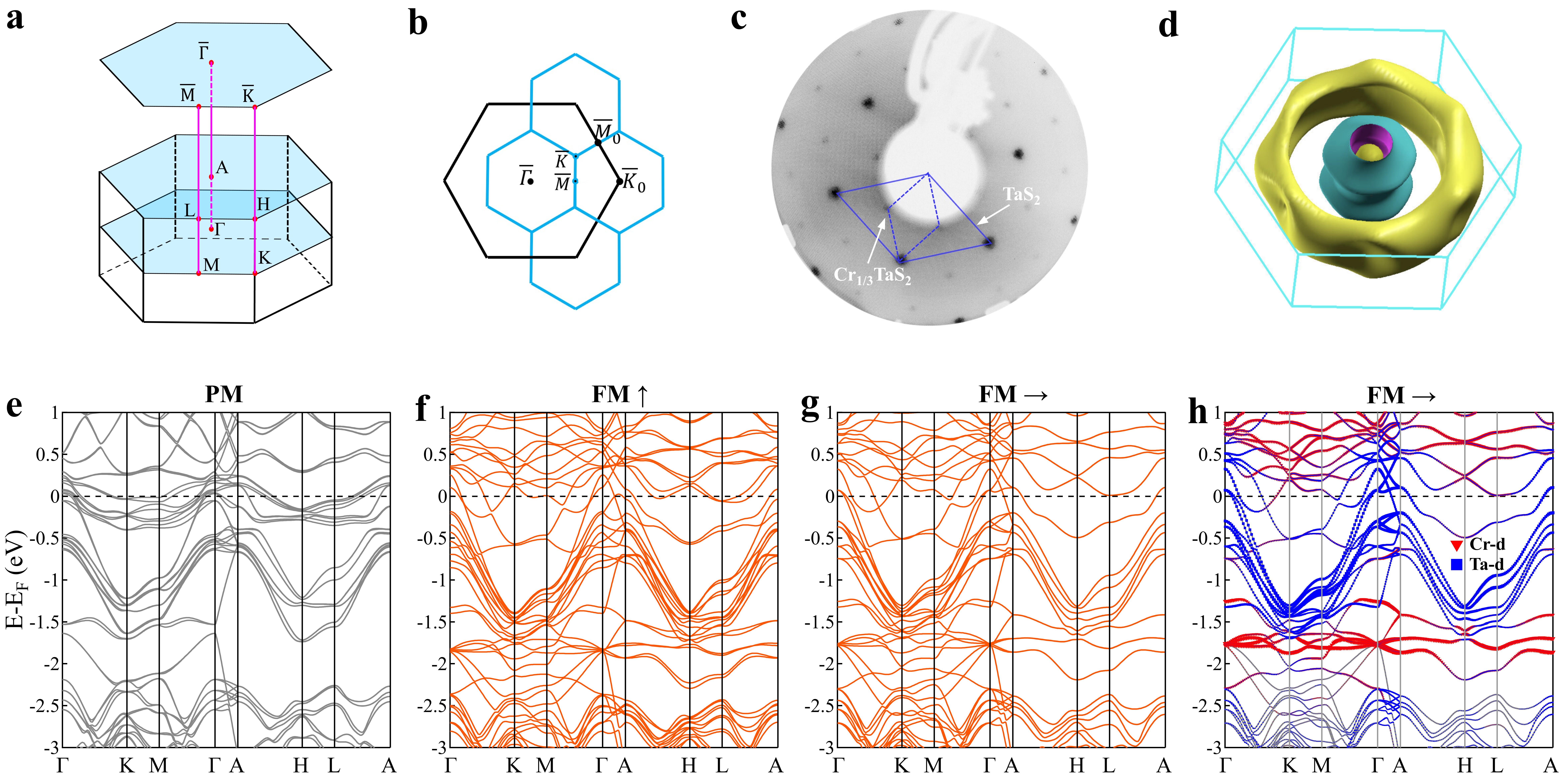}
  \caption{Bulk electronic structure of Cr$_{1/3}$TaS$_2$. (a) Schematic plot of the 3D bulk Brillouin zone and its projection onto the (001) surface Brillouin zone. (b) Surface Brillouin zones of the primitive $1\times1$ 2H-TaS$_2$ (black hexagon) and the $\sqrt{3}\,\times\sqrt{3}$ superlattice of Cr$_{1/3}$TaS$_2$ (blue hexagon). (c) LEED pattern of Cr$_{1/3}$TaS$_2$ measured at 72\,K with an electron beam energy of 220\,eV. The reciprocal lattices corresponding to the $1\times1$ (2H-TaS$_2$) and $\sqrt{3}\,\times\sqrt{3}$ (Cr$_{1/3}$TaS$_2$) periodicities are marked by solid lines and dashed lines, respectively. (d) Calculated bulk Fermi surface of Cr$_{1/3}$TaS$_2$ for the in-plane FM configuration. (e) Bulk band structure calculated for the paramagnetic (PM) phase. (f,g) Bulk band structures calculated for the out-of-plane FM order (f) and the in-plane FM order (g), respectively, illustrating the magnetic-orientation-dependent band evolution. (h) Orbital-projected bulk band structure calculated for the in-plane FM order, highlighting the contributions from Cr-$d$ orbitals (red triangles) and Ta-$d$ orbitals (blue squares).}
  \label{Bulkband}
\end{figure}

The bulk electronic structure of Cr$_{1/3}$TaS$_2$ was studied using density functional theory (DFT) calculations. Figure \ref{Bulkband}a shows the 3D bulk Brillouin zone and its projection onto the (001) surface Brillouin zone, while Figure \ref{Bulkband}b compares the surface Brillouin zones of the primitive $1\times1$ 2H-TaS$_2$ lattice and the $\sqrt{3}\,\times\sqrt{3}$ superlattice of Cr$_{1/3}$TaS$_2$ induced by Cr intercalation. Correspondingly, low-energy electron diffraction (LEED) measurements (Figure \ref{Bulkband}c) display sharp diffraction spots associated with both lattices, confirming the well-defined long-range superlattice periodicity. The calculated bulk Fermi surface (Figure \ref{Bulkband}d) under an in-plane FM order exhibits only Fermi pockets around the center of the superlattice Brillouin zone, with no additional electronic features appearing at the Brillouin-zone boundary. Notably, to facilitate direct comparison with our experimental results, all calculations presented in this work were performed within an effective ferromagnetic approximation, rather than explicitly modeling the CHM state, which would require prohibitively large supercells containing hundreds to thousands of atoms. We consider this approximation to be well justified: vacuum ultraviolet (VUV) ARPES, performed with photon energies of $h\nu$ = 48, 79, and 120\,eV in this study, is an extremely surface-sensitive probe, detecting photoelectrons emitted from only the topmost $\sim$10\,$\text{\AA}$ beneath the crystal surface.\autocite{Electronescapedepth} Compared with the long periodicity of CHM ($\sim$16\,nm) in Cr$_{1/3}$TaS$_2$, this probing depth spans only 1--2 adjacent Cr magnetic layers, within which the relative in-plane spin canting is minimal. This consideration is further supported by our subsequent investigation of the termination-dependent electronic structure.

The bulk band structure in the paramagnetic (PM) phase (Figure \ref{Bulkband}e) features multiple dispersive bands crossing the Fermi level. In contrast, once ferromagnetic order is included, the electronic structure becomes strongly spin-split (Figure \ref{Bulkband}f,g). More importantly, the out-of-plane and in-plane FM orders give rise to distinct band degeneracies at the $k_z = \pi/c$ plane (A--H--L--A high-symmetry path), which can be traced back to the different magnetic space-group symmetries associated with magnetization directions. Specifically, the band degeneracy on this plane is protected by an antiunitary combined symmetry $\{C_{2z}\mid 0,0,\frac{1}{2}\}T$, where $\{C_{2z}\mid 0,0,\frac{1}{2}\}$ denotes a twofold screw rotation and $T$ represents time-reversal symmetry. Owing to the antiunitary nature and $(\{C_{2z}\mid 0,0,\frac{1}{2}\}T)^2 =-1$ on the $k_z = \pi/c$ plane, a Kramers-like twofold degeneracy is enforced throughout that plane. When the magnetic moments align within the ab plane, this symmetry is preserved and the band degeneracy at $k_z = \pi/c$ remains intact. In contrast, for out-of-plane magnetization, the symmetry is broken, resulting in a lifting of the symmetry-protected band degeneracy at the $k_z = \pi/c$ plane. Furthermore, our orbital-resolved calculations (Figure \ref{Bulkband}h) further reveal substantial hybridization between Cr-3$d$ and Ta-5$d$ orbitals at low energies, indicating strong coupling between the intercalated magnetic moments and itinerant electrons in the Cr$_{1/3}$TaS$_2$. Together, these results establish a solid theoretical foundation for understanding the termination-dependent surface electronic structures discussed below.

\subsection{Termination-dependent surface electronic structure}

Due to the van der Waals-layered structure of Cr$_{1/3}$TaS$_2$, two distinct surface terminations can be naturally obtained upon cleavage: a TaS$_2$-terminated surface and a Cr-terminated surface. These two terminations host strikingly different electronic structures, as revealed by our high-resolution micro-ARPES and surface-state-resolved DFT calculations, as discussed below.

\begin{figure}
  \includegraphics[width=\linewidth]{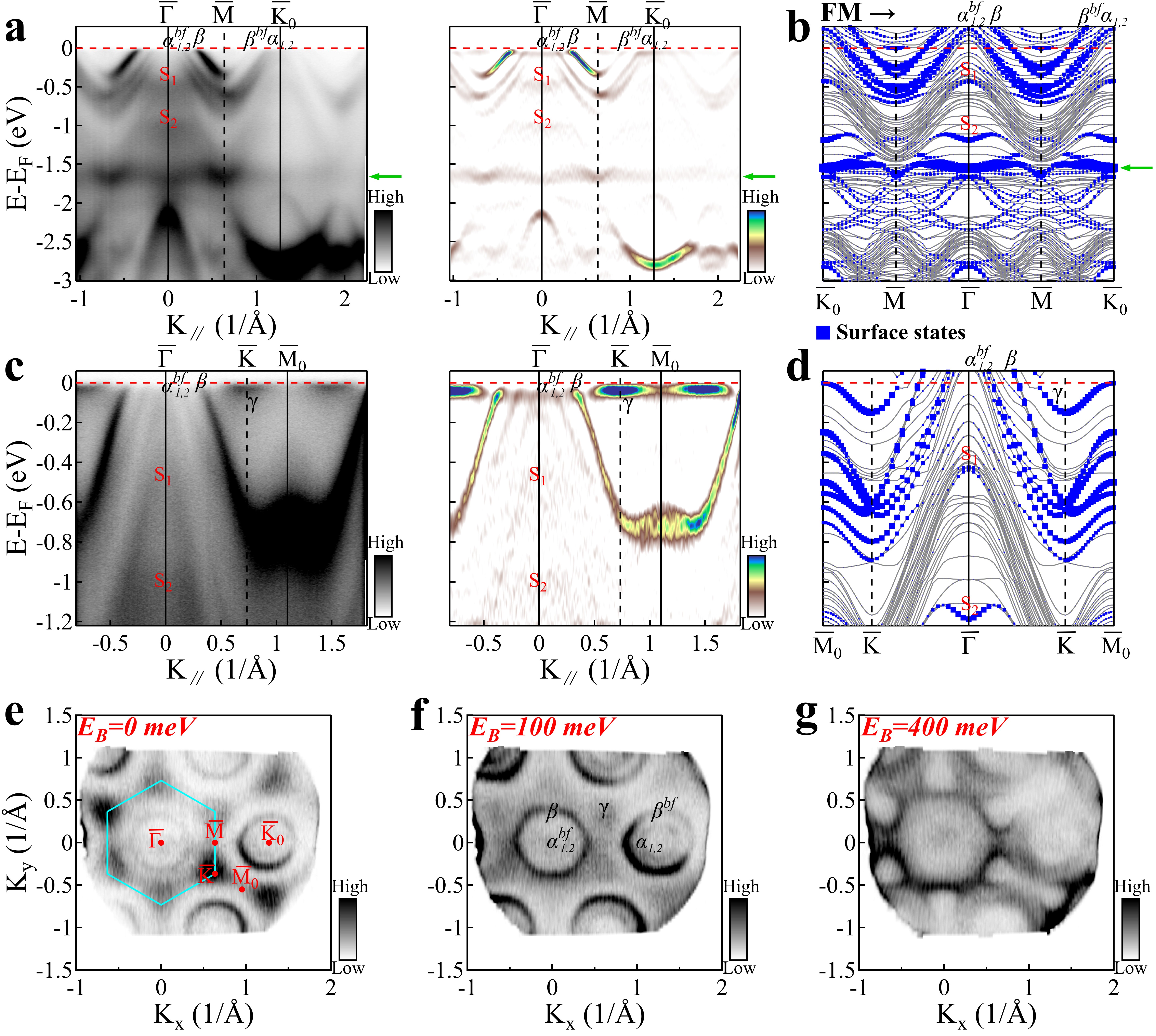}
  \caption{Electronic structure of the TaS$_2$-terminated surface. (a) (Left to right) ARPES spectra measured along the $\bar{\Gamma}$--$\bar{M}$--$\bar{K}_0$ high symmetry direction and the corresponding second-derivative image with respect to energy. The data in (a) were measured using synchrotron-based ARPES with a photon energy of 120\,eV at 10\,K. (b) Calculated surface band structures along the $\bar{\Gamma}$--$\bar{M}$--$\bar{K}_0$ path for the in-plane FM order, with surface states highlighted by blue squares. The flat-band position is indicated by the green arrows in (a,b). (c) (Left to right) ARPES spectra along the $\bar{\Gamma}$--$\bar{K}$--$\bar{M}_0$ high symmetry direction and the corresponding second-derivative image with respect to energy. (d) Calculated surface band structures along $\bar{\Gamma}$--$\bar{K}$--$\bar{M}_0$ path for the in-plane FM order, with surface states highlighted by blue squares. Note that the energy scales in (a,b) and (c,d) are different. (e--g) Fermi surface (e) and constant-energy contours at binding energy of 100\,meV (f) and 400\,meV (g), obtained by integrating the spectral intensity within 10\,meV with respect to the energy. The data in (c--g) were acquired using synchrotron-based ARPES with a photon energy of 79\,eV at 10\,K.}
  \label{TaS2term}
\end{figure}

Figure \ref{TaS2term}a,c present ARPES spectra measured along the $\bar{\Gamma}$--$\bar{M}$--$\bar{K}_0$ and $\bar{\Gamma}$--$\bar{K}$--$\bar{M}_0$ high-symmetry directions on the TaS$_2$-terminated surface, where several highly dispersive Ta-derived bands crossing the Fermi level are clearly resolved (see Figure S1, Supporting Information, for detailed orbital-projected calculations). The overall band features closely resemble the folded band structure of monolayer (ML) 2H-TaS$_2$\autocite{MLTaS2PhysRevB,MLTaS2ACSnano} subject to a $\sqrt{3}\,\times\sqrt{3}$ superlattice potential (see Figure S3, Supporting Information, for the calculated band structure of monolayer 2H-TaS$_2$ and the corresponding band-folding simulations): the original monolayer-like bands are labeled as $\alpha_{1,2}$ and $\beta$, while their backfolded counterparts are denoted as $\alpha_{1,2}^{bf}$ and $\beta^{bf}$, arising from the folding of $\bar{K}_0$ onto $\bar{\Gamma}$ and $\bar{\Gamma}$ onto $\bar{K}_0$. Notably, the observed SOC-induced band splitting near the $\bar{K}_0$ and $\bar{K}_0'$ points reaches approximately 170\,meV (Figure S4, Supporting Information), in good agreement with our DFT calculations (Figure S3, Supporting Information) as well as previous ARPES measurements on ML 2H-TaS$_2$,\autocite{MLTaS2PhysRevB,MLTaS2ACSnano} further corroborating the monolayer-like nature of these surface bands. In contrast to the ML 2H-TaS$_2$, however, the presence of the $\sqrt{3}\,\times\sqrt{3}$ superlattice potential in Cr$_{1/3}$TaS$_2$ leads to a pronounced hybridization between the original and backfolded bands, opening a sizable gap of $\sim$260\,meV at the $\bar{M}$ point (Figure S5, Supporting Information). Surface-state calculations for the TaS$_2$-terminated surface (Figure \ref{TaS2term}b,d; Figure S1, Supporting Information) further predict the presence of monolayer-like bands and their backfolded counterparts ($\alpha_{1,2}$ and $\beta$, $\alpha_{1,2}^{bf}$ and $\beta^{bf}$), together with two additional surface-derived parabolic bands, namely the hole-like band $S_1$ and the electron-like band $S_2$, originating predominantly from the intercalated Cr atoms. Importantly, the calculations faithfully reproduce the hybridization between the monolayer-like TaS$_2$ surface states and their backfolded counterparts at the $\bar{M}$ point, resulting in a hybridization gap (highlighted by orange arrows in calculations, Figure S1b, Supporting Information) consistent with the experimental observations. We note that the calculated band splitting of the $\alpha_{1,2}$ bands near the $\bar{K}_0$ point is relatively weak. This may stem from the underestimation of SOC effect on the surface within theoretical calculations. To test this scenario, we present additional calculations with different enhanced SOC strengths in Figure S6 (Supporting Information). Indeed, the band splitting near the $\bar{K}_0$ point increases significantly with increasing SOC strength.

Furthermore, the pronounced spectral-weight enhancement observed experimentally along the $\bar{K}$--$\bar{M}_0$ path at a binding energy of $\sim$0.8\,eV (Figure \ref{TaS2term}c) corresponds to the convergence of multiple split surface-state bands in the calculations (Figure \ref{TaS2term}d), reflecting a fundamental symmetry difference between the chiral helimagnet Cr$_{1/3}$TaS$_2$ and ML 2H-TaS$_2$. In ML 2H-TaS$_2$, the $\bar{M}_0$ point is a time-reversal-invariant momentum (TRIM) point, where band degeneracy is strictly protected by time-reversal symmetry. Furthermore, mirror symmetry enforces this degeneracy along the entire $\bar{\Gamma}$--$\bar{M}_0$ high-symmetry line. By contrast, the Cr$_{1/3}$TaS$_2$ slab structure belongs to the $C_3$ point group, and the emergence of magnetic order breaks time-reversal symmetry. This symmetry reduction lifts the band degeneracy at the $\bar{M}_0$ point and along $\bar{\Gamma}$--$\bar{M}_0$ line, resulting in a broad energy distribution of $\sim$500\,meV around $\bar{M}_0$ that is consistently manifested in both experiment and theory.

Monolayer-like TaS$_2$ surface states are reminiscent of the V-intercalated transition metal dichalcogenide V$_{1/3}$NbS$_2$,\autocite{VNSPDCKing} where monolayer-like band features were attributed to surface charge redistribution induced by cleaving intercalated TMD materials. According to the polar catastrophe model,\autocite{HaroldHwang} the TMD-terminated surface becomes hole-doped relative to the bulk, allowing the surface states to be energetically separated from the bulk bands. Although both exhibit monolayer-like TMD electronic band structures, an important distinction emerges when comparing Cr$_{1/3}$TaS$_2$ with V$_{1/3}$NbS$_2$ in terms of valley-dependent band splitting near the $\bar{K}_0$ and $\bar{K}_0'$ points of the monolayer-like TMD bands. In V$_{1/3}$NbS$_2$, the presence of an out-of-plane magnetic component gives rise to an additional valley--Zeeman splitting that superimposes on the intrinsic valley spin--orbit coupling--induced band splitting, whose sign is opposite at $\bar{K}_0$ and $\bar{K}_0'$. As a result, the band splitting around $\bar{K}_0$ and $\bar{K}_0'$ exhibits an antisymmetric behavior, with the splitting enhanced at one valley while being suppressed at the symmetry-related opposite momentum. By contrast, such valley-dependent antisymmetric band splitting is not observed in Cr$_{1/3}$TaS$_2$. Instead, the band splittings near $\bar{K}_0$ and $\bar{K}_0'$ are found to be symmetric and remain nearly temperature independent (Figure S4, Supporting Information). This behavior reflects the intrinsically unusual character of the chiral magnetic order in Cr$_{1/3}$TaS$_2$, where the magnetic moments within each layer are aligned collinearly in the in-plane direction, while their orientations rotate continuously along the crystallographic $c$ axis between adjacent layers, resulting in the absence of a net out-of-plane magnetic component. The symmetric valley band splitting in Cr$_{1/3}$TaS$_2$, together with the magnetic-field tunability of its chiral magnetic states, provides a promising route toward efficient and low-power valleytronic control. A small magnetic field applied along the $c$ axis ($H \parallel c$) can stabilize a chiral conical phase, while a tilted field drives the system into a tilted chiral soliton lattice,\autocite{PhysRevB.96.184423} both of which feature spin moments with a sizable out-of-plane component. This enables a controllable crossover from symmetric to asymmetric valley-dependent electronic behavior and an effective tuning of the valley band splitting around the $\bar{K}_0$ and $\bar{K}_0'$ points. Such control of valley--spin splittings via exchange coupling between intrinsic localized magnetic moments and itinerant electrons is expected to be far more efficient than approaches based on direct magnetic-field--induced Zeeman splitting\autocite{srivastava2015valley} or heterostructure-based proximity coupling.\autocite{zhong2017van,norden2019giant}

The Fermi surface and constant-energy contours (CECs) (Figure \ref{TaS2term}e--g) provide further evidence for the superlattice-modified Fermiology. While the overall Fermi surface retains a hexagonal topology characteristic of pristine 2H-TaS$_2$, additional replicas and pronounced intensity modulations emerge due to the Cr-induced $\sqrt{3}\,\times\sqrt{3}$ periodic potential. In particular, a shallow electron pocket $\gamma$ near the Fermi level appears at the $\bar{K}$ point of the superlattice Brillouin zone (Figure \ref{TaS2term}c--f), forming a characteristic triangular Fermi surface and CEC geometry---an electronic feature absent in pristine 2H-TaS$_2$ (Figure S3d, Supporting Information). This behavior also stands in sharp contrast to the bulk Fermi surface of Cr$_{1/3}$TaS$_2$ calculated under in-plane FM order (Figure \ref{Bulkband}d), where no associated electronic features are present at the superlattice Brillouin-zone boundary. Orbital-projected surface-state calculations (see Figure S1c,d, Supporting Information) further reveal that the triangular $\gamma$ pocket originates from hybridization between Cr-3$d$ and Ta-5$d$ orbitals. Similar triangular shallow electron pockets have previously been reported in Co$_{1/3}$NbS$_2$\autocite{CoNSKondo,CoNSHasan,CoNSDMFT} and Co$_{1/3}$TaS$_2$,\autocite{luo2026CoTaS} two magnetic-intercalated TMD compounds exhibiting anomalous Hall effect but lacking chiral helimagnetic order, where they were attributed to electronic states derived from the magnetic Co intercalants. In contrast, for Cr$_{1/3}$TaS$_2$, our combined experimental and theoretical results unambiguously demonstrate that the triangular $\gamma$ pocket originates from Cr--Ta hybridization, highlighting the unusual coupling between the localized Cr moments and the itinerant electrons in chiral helimagnet Cr$_{1/3}$TaS$_2$. This coupling will be further elucidated in our temperature-dependent investigation of the surface electronic structure.

\begin{figure}
  \includegraphics[width=\linewidth]{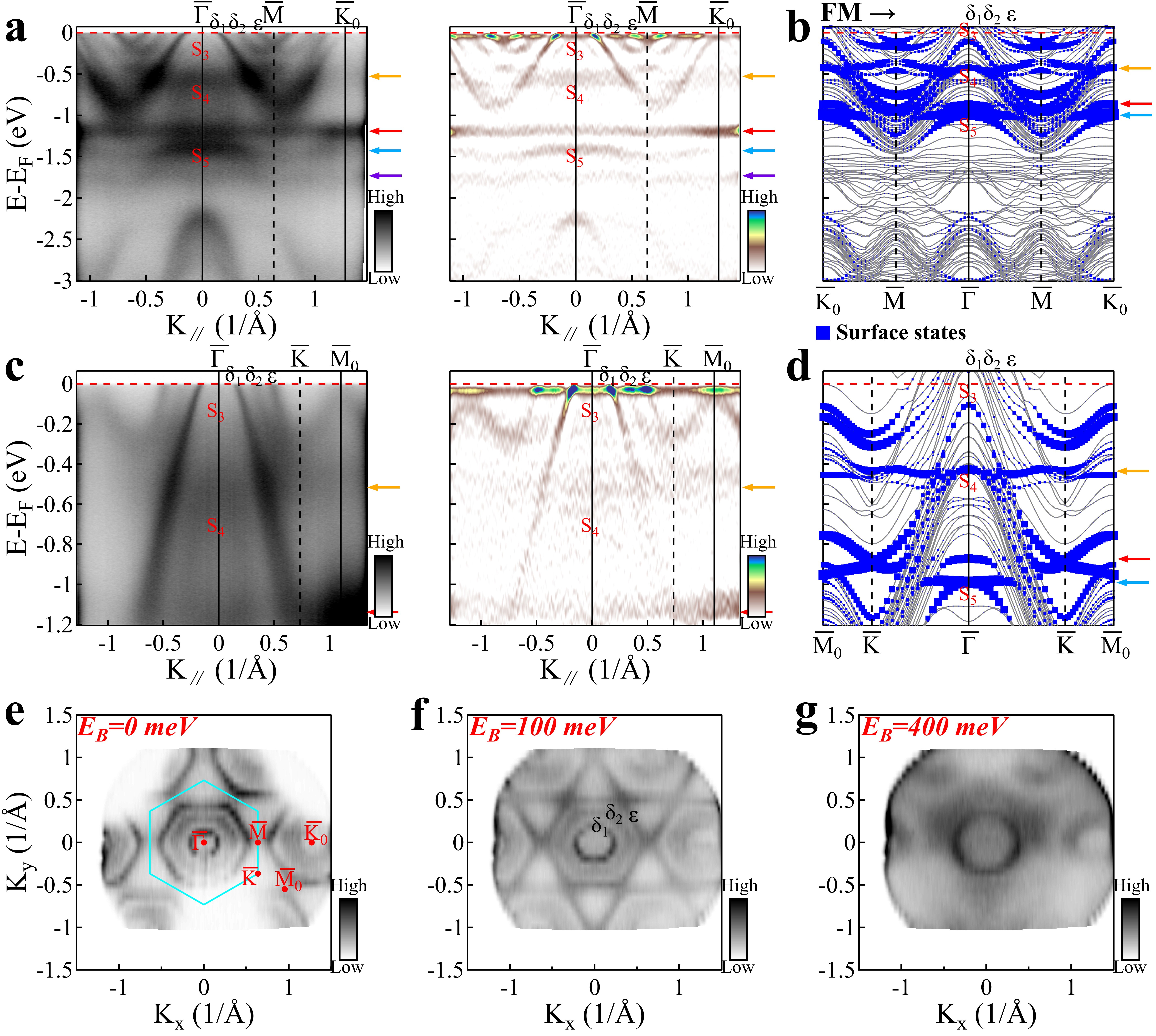}
  \caption{Electronic structure of the Cr-terminated surface. (a) (Left to right) ARPES spectra measured along the $\bar{\Gamma}$--$\bar{M}$--$\bar{K}_0$ high symmetry direction and the corresponding second-derivative image with respect to energy. Four distinct flat-band features are observed and highlighted by arrows in orange, red, blue, and purple, respectively. (b) Calculated surface band structures along the $\bar{\Gamma}$--$\bar{M}$--$\bar{K}_0$ path for the in-plane FM order, with surface states highlighted by blue squares. (c) (Left to right) ARPES spectra measured along the $\bar{\Gamma}$--$\bar{K}$--$\bar{M}_0$ high symmetry direction and the corresponding second-derivative image with respect to energy. (d) Calculated surface band structures along the $\bar{\Gamma}$--$\bar{K}$--$\bar{M}_0$ path for the in-plane FM order, with surface states highlighted by blue squares. Note that the energy scales in (a,b) and (c,d) are different. (e--g) Fermi surface (e) and constant-energy contours at binding energy of 100\,meV (f) and 400\,meV (g), obtained by integrating the spectral intensity within 10\,meV with respect to the energy. The data shown in a,c, and e-g were acquired at 10\,K using synchrotron-based ARPES with a photon energy of 79\,eV.}
  \label{Crterm}
\end{figure}

Unlike the TaS$_2$-terminated surface, where the folded monolayer-like TaS$_2$ bands emerge and cross the Fermi level, the Cr-terminated surface does not exhibit such features. Instead, three distinct hole-like pockets, $\delta_1$, $\delta_2$, and $\varepsilon$, are observed around the $\bar{\Gamma}$ point (Figure \ref{Crterm}a,c,f). Surface-state calculations accurately reproduce these band features (Figure \ref{Crterm}b,d; Figure S2, Supporting Information), and further reveal additional hole-like parabolic surface bands, denoted as $S_3$, $S_4$, and $S_5$. To the best of our knowledge, these constitute the first direct experimental confirmation of a definitive surface electronic band-structure assignment on a magnetic transition-metal-ion-terminated surface in the family of intercalated TMDs. By contrast, previous studies on Cr$_{1/3}$NbS$_2$\autocite{CNS2016PhysRevB} and Co$_{1/3}$TaS$_2$\autocite{luo2026CoTaS} suggested that the magnetic intercalant atoms are disordered on the cleaved sample surface, making it difficult to form a well-defined periodic structure and thereby hindering the identification of their intrinsic electronic characteristics. Corresponding constant-energy contours (Figure \ref{Crterm}e--g) reveal the same hole-pocket features $\delta_1$, $\delta_2$, and $\varepsilon$ around both the $\bar{\Gamma}$ and $\bar{K}_0$ points, demonstrating a peculiar reconstruction of the surface electronic structure induced by the intercalated Cr-ordered lattice. Interestingly, the $\varepsilon$ pocket further evolves at a binding energy of $\sim$100\,meV into a momentum-space kagome network. While orbital-projected calculations show that, in sharp contrast to the TaS$_2$-terminated surface---where the high--Fermi-velocity conduction bands near $E_F$ are predominantly of Ta $d$-orbital character---the Cr-terminated surface introduces a substantially enhanced contribution from Cr-$d$ orbitals near the Fermi level through hybridization (see Figure S2, Supporting Information).

Taken together, the remarkable hybridization and reconstruction of the low-energy electronic structure observed on both the TaS$_2$- and Cr-terminated surfaces of Cr$_{1/3}$TaS$_2$ highlight the profound impact of Cr intercalation on the electronic properties. This finding is further supported by previous ResPES studies on the sister compound Cr$_{1/3}$NbS$_2$,\autocite{cNS2020CP} which revealed a Fermi surface composed of strongly hybridized Cr- and Nb-derived electronic states.
Our termination-resolved measurements led to new findings, that the extent and symmetry of the low-energy hybridization---and the resulting Fermi surface topology---can be strongly termination dependent. Therefore, the corresponding  ARPES features actually cannot be captured merely by bulk electronic structure descriptions, which were done almost without exception by previous ARPES studies of chiral helimagnetic materials.\autocite{CNS2016PhysRevB,cNS2020CP,CYLCNSPRB,XieLLARPES}

In addition, we observe multiple distinct flat-band features with extremely narrow bandwidths on both the TaS$_2$- and Cr-terminated surfaces. On the TaS$_2$-terminated surface, a prominent flat band appears at $\sim$1.648\,$\pm$\,0.042\,eV below the $E_F$ (green arrow in Figure \ref{TaS2term}a; see also Figure S7a,c, Supporting Information for details), which is also reproduced by surface-state calculations (Figure \ref{TaS2term}b; see also Figure S1, Supporting Information). On the Cr-terminated surface, four flat bands are observed at binding energies of $\sim$0.535\,$\pm$\,0.035\,eV, $\sim$1.183\,$\pm$\,0.013\,eV, $\sim$1.423\,$\pm$\,0.043\,eV, and $\sim$1.735\,$\pm$\,0.021\,eV (orange, red, blue, and purple arrows in Figure \ref{Crterm}a,c; see also Figure S7b,d, Supporting Information for details). Although our surface-state calculations for the topmost Cr-terminated surface yield only three flat-band features (Figure \ref{Crterm}b,d; see also Figure S2, Supporting Information), inclusion of electronic contributions from the subsurface layer reproduces four flat bands very well (Figure S8, Supporting Information), as well as a pronounced spectral-weight enhancement of the $\delta_1$ band, both in excellent agreement with experiment. Orbital-resolved analyses (see Figure S1 and S2, Supporting Information) further confirm that all flat bands on the TaS$_2$- and Cr-terminated surfaces originate from Cr-$d$ orbitals, underscoring the strong influence of intercalated Cr on the local electronic environment. With these flat bands clearly appearing at the same binding energies along both the $\bar{\Gamma}$--$\bar{M}$--$\bar{K}_0$ and $\bar{\Gamma}$--$\bar{K}$--$\bar{M}_0$ directions, which may indicate a nearly uniform spectral weight distribution across the entire 2D Brillouin zone. 

Regarding the origin of flat bands in Cr$_{1/3}$TaS$_2$, a recent theoretical study has suggested that they may arise from an unconventional interlayer geometric frustration,\autocite{peng2025flat} in which destructive quantum interference between hopping processes from the intercalated transition-metal atoms to the neighboring S atoms and those from Ta atoms to the same S sites leads to strong wave-function localization within the Cr-S-Ta trigonal bipyramidal structures. In the present system, the distinct energy positions of the flat bands observed on different surface terminations are therefore most likely attributable to variations of the surface electrostatic potential. The excellent agreement between theory and experiment for the extremely narrow surface flat bands further supports the surface origin of the observed electronic structure in Cr$_{1/3}$TaS$_2$, and establishes Cr$_{1/3}$TaS$_2$ as a new materials platform for exploring the flat-band--driven correlation and topological phenomena.\autocite{Yin2022Kagomereview,Nuckolls2024}

\subsection{Unusual temperature evolution of electronic states and magnetic band splitting on distinct terminations}

\begin{figure}
  \includegraphics[width=\linewidth]{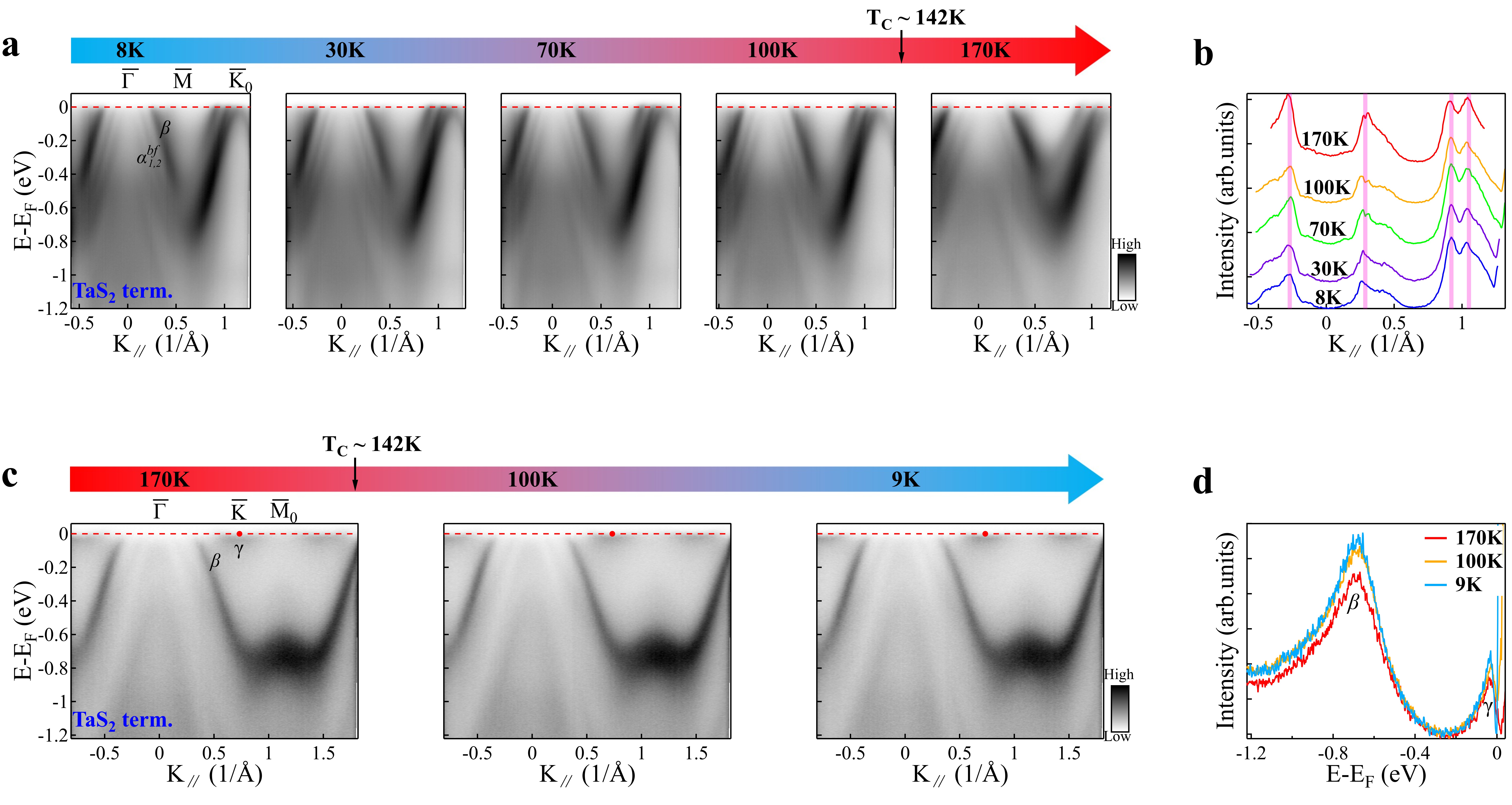}
  \caption{Temperature evolution of the electronic structure on the TaS$_2$-terminated surface. (a) Temperature-dependent band structures measured along the $\bar{\Gamma}$--$\bar{M}$--$\bar{K}_0$ high-symmetry direction on the TaS$_2$-terminated surface as the temperature increases from 8\,K to 170\,K, crossing the magnetic transition temperature $T_C$$\sim$142\,K. The data in (a) were acquired by using synchrotron-based ARPES with a photon energy of 48\,eV. (b) MDCs at the Fermi level extracted from the data in (a). (c) Temperature evolution of the band dispersions measured along the $\bar{\Gamma}$--$\bar{K}$--$\bar{M}_0$ high-symmetry path on the TaS$_2$-terminated surface as the temperature decreases from 170\,K to 9\,K, acquired with a photon energy of 79\,eV. (d) EDCs at the $\bar{K}$ point, as indicated by the red dots in (c). The data have been divided by the Fermi--Dirac distribution.}
  \label{TaS2termtemperature}
\end{figure}

Beyond the pronounced difference in the electronic structures of the two surface terminations, their surface states also exhibit distinctly different temperature evolutions. Figure \ref{TaS2termtemperature} summarizes the temperature-dependent ARPES measurements performed on the TaS$_2$-terminated surface. Based on the analysis presented above, the unique CHM spin texture in Cr$_{1/3}$TaS$_2$, which lacks an out-of-plane magnetization component, is not expected to induce a sizable change in the SOC-driven band splitting at the $\bar{K}_0$ and $\bar{K}_0'$ valleys through magnetic exchange coupling between the intercalated Cr layer and the surface TaS$_2$ layer (Figure S4, Supporting Information). This naturally raises two important questions: (i) whether other magnetically split surface bands exist on the TaS$_2$-terminated surface; and (ii) whether the triangular shallow electron pocket $\gamma$, arising from Cr-3$d$--Ta-5$d$ orbital hybridization, persists above the magnetic transition temperature. These issues call for a systematic investigation of the temperature evolution of the surface electronic structure. Along the $\bar{\Gamma}$--$\bar{M}$--$\bar{K}_0$ direction, the band dispersions show no discernible changes as the temperature is increased from 8\,K to 170\,K (Figure \ref{TaS2termtemperature}a). This observation is further substantiated by a comparison of momentum distribution curves (MDCs) extracted at the Fermi level (Figure \ref{TaS2termtemperature}b), where the peak positions ($k_F$) associated with the relevant band features remain essentially unchanged across the magnetic transition temperature ($T_C$$\sim$142\,K).

Upon cooling, we further investigated the temperature evolution of the electronic structure along the $\bar{\Gamma}$--$\bar{K}$--$\bar{M}_0$ direction (Figure \ref{TaS2termtemperature}c). Similar to the $\bar{\Gamma}$--$\bar{M}$--$\bar{K}_0$ cut, no apparent changes are observed across $T_C$, including for the shallow electron pocket $\gamma$ located at the superlattice Brillouin-zone $\bar{K}$ point. Although the TaS$_2$-terminated surface does not display a resolvable magnetic band splitting across the phase transition, a more detailed comparison of normalized energy distribution curves (EDCs) taken at the central position of the $\gamma$ pocket reveals a clear temperature dependence of the peak intensities for the $\gamma$ and $\beta$ bands (Figure \ref{TaS2termtemperature}d). From 9\,K to 100\,K, their peak intensity remains nearly unchanged, whereas a clear reduction appears at 170\,K, well above $T_C$. Notably, this intensity evolution closely follows the temperature dependence of the magnetization ($M-T$) curve (Figure \ref{Lattice}f). Combined with orbital-projected calculations (Figure S1c,d, Supporting Information) showing hybridization between Cr-3$d$ and Ta-5$d$ orbitals, these results indicate that the TaS$_2$-terminated surface states are still coupled to the underlying Cr magnetic order, despite the absence of an observable magnetic band splitting. Similar temperature-dependent intensity variations consistent with the magnetic transition have also been reported for the $\gamma$-like electronic states in the chiral antiferromagnet Co$_{1/3}$TaS$_2$,\autocite{xiong2025all} supporting the notion that such intensity evolution reflects electron coupling to the underlying magnetic order.

\begin{figure}
  \includegraphics[width=\linewidth]{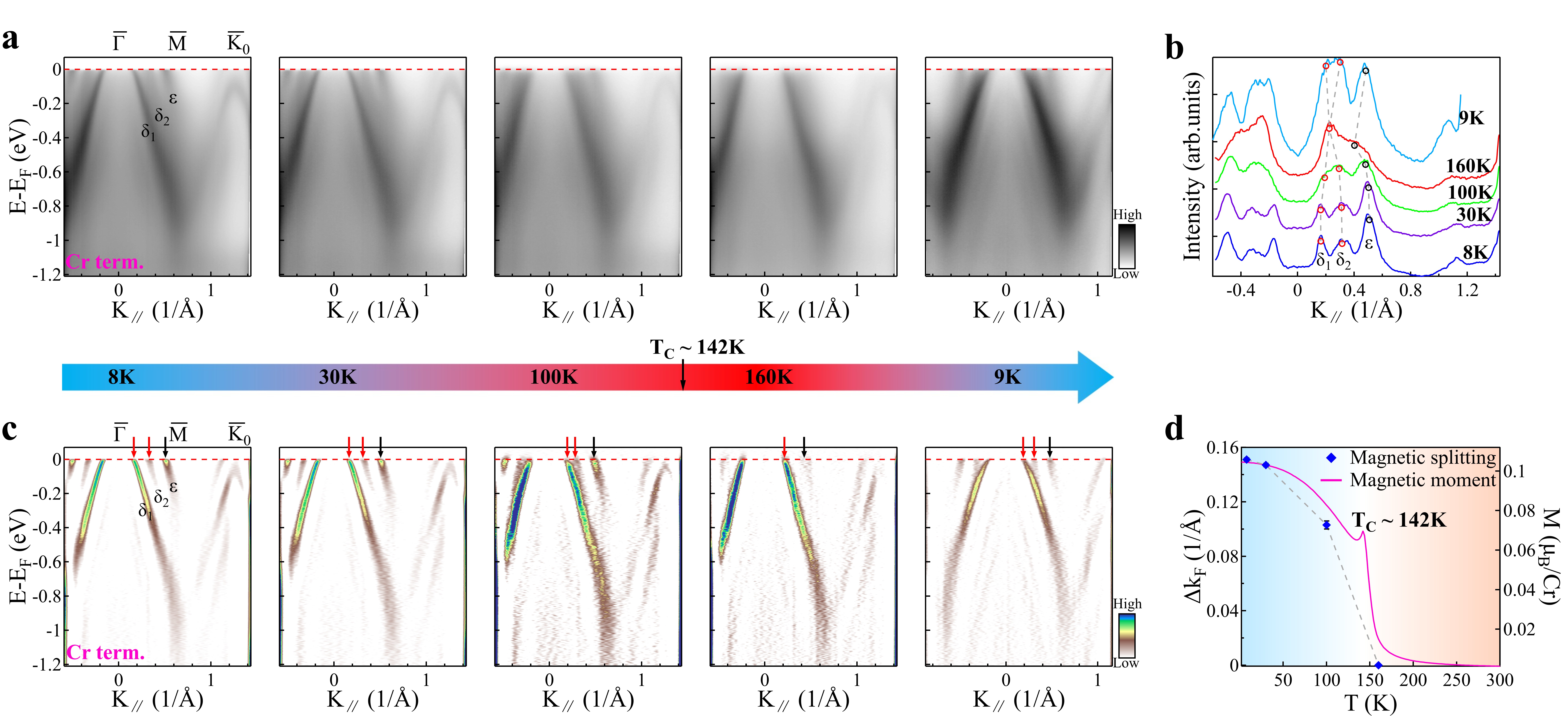}
  \caption{Magnetic band splitting on the Cr-terminated surface. (a) Temperature evolution of the electronic bands measured along the $\bar{\Gamma}$--$\bar{M}$--$\bar{K}_0$ high-symmetry direction on the Cr-terminated surface, as the temperature increases from 8\,K to 160\,K and subsequently returns to 9\,K. The data in (a) were acquired using synchrotron-based ARPES with a photon energy of 48\,eV. (b) MDCs at the Fermi level extracted from (a). The peak positions of $\delta_1$, $\delta_2$, and $\varepsilon$ bands are marked by red and black circles, respectively. (c) Second-derivative images with respect to momentum obtained from the data in (a). The momentum positions of $\delta_1$, $\delta_2$, and $\varepsilon$ bands are marked by red and black arrows, respectively. (d) Temperature dependence of the magnetic band splitting (blue diamonds) extracted from the ARPES measurements in (a), plotted together with the magnetization $M-T$ curve (pink line).}
  \label{CrtermTemperature}
\end{figure}

In stark contrast, the Cr-terminated surface exhibits pronounced magnetic band splitting with a strong temperature dependence. The ARPES spectra along the $\bar{\Gamma}$--$\bar{M}$--$\bar{K}_0$ direction (Figure \ref{CrtermTemperature}a) show three prominent bands ($\delta_1$, $\delta_2$, and $\varepsilon$) whose momentum positions shift markedly and are accompanied by a clear band splitting as the temperature crosses $T_C$. The corresponding second-derivative images with respect to momentum clearly highlight the splitting between the $\delta_1$ and $\delta_2$ branches (red arrows in Figure \ref{CrtermTemperature}c). A quantitative analysis based on MDCs extracted at the Fermi level (Figure \ref{CrtermTemperature}b) reveals that, upon warming, the $\delta_1$ and $\delta_2$ bands progressively approach each other, resulting in a continuous reduction of the splitting, while the $\varepsilon$ band remains largely unaffected. Once the temperature exceeds $T_C$$\sim$142\,K, the splitting between $\delta_1$ and $\delta_2$ disappears entirely, accompanied by an abrupt momentum shift of the $\varepsilon$ band, a behavior that has also been reported in Cr$_{1/3}$NbS$_2$.\autocite{CYLCNSPRB} The extracted splitting amplitude $\Delta k_F$ at the Fermi level (blue diamonds in Figure \ref{CrtermTemperature}d) decreases sharply near $T_C$ and closely tracks the magnetization curve (pink line), further corroborating the magnetic origin of the splitting. 
This quantitative correspondence establishes the band splitting between the $\delta_1$ and $\delta_2$ surface states as an electronic fingerprint of the CHM in Cr$_{1/3}$TaS$_2$. The surface-state characteristics on Cr-terminated surface are therefore highly sensitive to the underlying magnetic order, establishing ARPES as a direct spectroscopic probe of the CHM order parameter. 

Together, the pronounced temperature-dependent evolution of the surface-state bands, combined with their clear Cr--Ta orbital hybridization character, reveals a strong exchange interaction between the intercalated localized Cr magnetic moments and the itinerant electrons, going beyond the conventional weak-coupling Ruderman--Kittel--Kasuya--Yosida (RKKY) mechanism, thereby positioning Cr$_{1/3}$TaS$_2$ as a promising platform for exploring magnetically correlated surface states with tunable spin-dependent electronic properties, as well as for potential applications in chiral-magnet--based spintronic devices. 

\section{Conclusion}

In summary, we have systematically mapped the electronic landscape of the nano-scale chiral helimagnet Cr$_{1/3}$TaS$_2$ by combining high-resolution micro-ARPES and surface-state-resolved DFT calculations. Two distinct cleavage terminations host fundamentally different surface states: the TaS$_2$-terminated surface hosts folded monolayer-like TaS$_2$ bands shaped by the $\sqrt{3}\,\times\sqrt{3}$ superlattice potential and a shallow triangular electron pocket at the corners of the superlattice Brillouin zone, arising from Cr--Ta hybridization, whereas the Cr-terminated surface exhibits unusually reconstructed hole pockets with a significant temperature-dependent magnetic splitting. The temperature-driven disappearance of this splitting above $T_C$ identifies it as a direct electronic fingerprint of the underlying nano-scale chiral helimagnetic order. Moreover, multiple ultranarrow Cr $d$-orbital--derived surface flat bands are resolved and quantitatively reproduced by surface-state calculations, revealing strong magnetic correlations confined to the near-surface region. 
Therefore, our results elevate monoaxial chiral helimagnets from the bulk magnetotransport platforms to the interface-engineerable quantum materials, in which surface termination acts as a decisive control knob for the magnetic exchange field and the correlated surface electronic structure. Our findings unveil a strong interplay between the intercalated Cr moments and the itinerant electrons at surfaces, establishing Cr$_{1/3}$TaS$_2$ as a promising platform for magnetically correlated surface states with tunable spin and valley responses, and as a potential building block for chiral-magnet--based spintronic micro/nanodevices.

\section{Methods}

\subsection{Growth of single crystals}

Cr$_{1/3}$TaS$_2$ single crystals were successfully synthesized using the chemical vapor transport method. High-purity Cr powder (99.95\%), Ta pieces (99.95\%), and S powders (99.999\%) were mixed in a molar ratio of 0.5\,:\,1\,:\,2 and sealed under high vacuum in a quartz tube (10\,cm in length) with 0.1\,g of I$_2$ as the transport agent. The sealed ampoule was first heated in a muffle furnace at 400\,$^\circ$C for 72\,h to pre-react sulfur and reduce internal pressure, followed by reaction at 850\,$^\circ$C for 400\,h before furnace cooling. Subsequently, the ampoule was transferred to a two-temperature-zone furnace for crystal growth. The growth and source zones were set to 980\,$^\circ$C and 800\,$^\circ$C, respectively, and held for 48\,h to purify the nucleation region. After reversing the temperature gradient and maintaining the system for 336\,h, the crystals were finally quenched to room temperature.

\subsection{ARPES measurements}

Synchrotron-based ARPES measurements were performed with micro-focused VUV light at the 03U beamline of Shanghai Synchrotron Radiation Facility (SSRF)\autocite{Yang2021} with a hemispherical electron energy analyzer DA30L (Scienta-Omicron). The energy and momentum resolutions were set to better than 20\,meV and 0.02\,$\text{\AA}^{-1}$. The light spot size was smaller than 20\,$\mu\mathrm{m}$. The preliminary ARPES measurements were carried out on the BL13U beamline of the National Synchrotron Radiation Laboratory (NSRL), the BL09U beamline of the SSRF, and the BL-9B of Research Institute for Synchrotron Radiation Science (HiSOR). All the samples were cleaved in situ at a low temperature of 15\,K and measured in ultra-high vacuum with a base pressure better than 6$\times$10$^{-11}$\,mbar. The Fermi level is referenced by measuring on a clean polycrystalline gold that is electrically connected to the sample.

\subsection{Electronic structure calculations}

Our calculations are performed using density functional theory (DFT) as implemented in the Vienna ab initio simulation package (VASP) code.\autocite{Kresse1993,Akresse,BKresse} The Perdew-Burke-Ernzerhof (PBE) exchange-correlation functional and the projector-augmented-wave (PAW) approach are used. Throughout the work, the cutoff energy is set to be 450\,eV for expanding the wave functions into plane-wave basis. On the basis of the equilibrium structure, the k mesh used is $8\times8\times4$ and $8\times8\times1$ for bulk and slab. To simulate experimental TaS$_2$-terminated and Cr-terminated surface, a thin-film slab consisting of three unit cells along the surface normal were constructed (see Figure S8c,f, Supporting Information). The lattice constants used in our calculation are a = b = 5.76\,$\text{\AA}$ and c = 12.22\,$\text{\AA}$, taken from the Materials Project database\autocite{materialproject} (ID: mp-1189011).

\section*{Author Contributions}

$^{\dagger}$B.L., X.L., C.L., and Z.W. contributed equally to this work.

\section*{Notes}

The authors declare no competing financial interest.

\section*{Acknowledgements}

This work is supported by the National Key Research and Development Program of China (Grants No. 2022YFA1403901 and 2023YFA1406103 by G.L., 2022YFA1402600 by Y.-C.L., 2023YFA1407300 by X.W. and 2022YFA1604200 by L. Z.), the National Natural Science Foundation of China (Grants No. 12488201 by X.J.Z., 12374066 by L.Z., 12274438 by Y.-C.L., 12574151, 12447103, and 12447101 by X.W), the start-up fund from Hefei National Laboratory. The authors thank the Shanghai Synchrotron Radiation Facility of BL03U (\url{https://cstr.cn/31124.02.SSRF.BL03U}) and BL09U (\url{https://cstr.cn/31124.02.SSRF.BL09U}) for the assistance on ARPES measurements. The authors thank the staff members of the ARPES System (\url{https://cstr.cn/31131.02.HLS.ARPES}) at the National Synchrotron Radiation Laboratory in Hefei (\url{https://cstr.cn/31131.02.HLS}), for providing technical support and assistance in data collection and analysis. Synchrotron ARPES measurements at HiSOR were performed under the approval of the Program Advisory Committee (Proposal Numbers: 22BG045 and 23AG023).

\section*{Supporting information}

Additional experimental and computational data, including orbital-resolved surface-state calculations for the TaS$_2$- and Cr-terminated surfaces, comparison with monolayer 2H-TaS$_2$ and $\sqrt{3}\,\times\sqrt{3}$ superlattice-folded bands, analysis of symmetric valley band splitting around $\bar{K}_0$ and $\bar{K}_0'$, hybridization-gap and SOC-dependent band-splitting analyses, surface flat-band features, subsurface-layer effects, and cross-check bulk and slab calculations using the Quantum ESPRESSO package (PDF).

\printbibliography

\newpage

\end{document}